\documentclass[aps,prb,twocolumn,superscriptaddress,unsortedaddress]{revtex4}
\usepackage{epsfig}
\usepackage{amsmath}

\newcommand{\sbar}{\bar{s}}

\newcommand{\be}{\begin{equation}}
\newcommand{\ee}{\end{equation}}
\newcommand{\bk}{{{\bf{k}}}}
\newcommand{\bq}{{{\bf{q}}}}

\newcommand{\bQ}{{{\bf{Q}}}}
\newcommand{\br}{{{\bf{r}}}}

\newcommand{\bR}{{{\bf{R}}}}

\newcommand{\bea}{\begin{eqnarray}}
\newcommand{\eea}{\end{eqnarray}}
\newcommand{\ra}{\rangle}
\newcommand{\la}{\langle}
\newcommand{\upa}{\uparrow}
\newcommand{\dna}{\downarrow}
\newcommand{\bS}{{\bf S}}

\newcommand{\dg}{{\dagger}}
\newcommand{\pdg}{{\phantom\dagger}}
\newcommand{\nn}{\nonumber}

\begin{document}

\title{Spiral order by disorder and lattice nematic order in a frustrated Heisenberg antiferromagnet
on the honeycomb lattice}
\author{A. Mulder}
\affiliation{Department of Physics, University of Toronto, Toronto, Ontario M5S 1A7, Canada}
\author{R. Ganesh}
\affiliation{Department of Physics, University of Toronto, Toronto, Ontario M5S 1A7, Canada}
\author{L. Capriotti}
\affiliation{Quantitative Strategies,
Investment Banking Division, Credit Suisse Group, Eleven Madison
Avenue, New York City, NY 10010-3086, USA}
\author{A. Paramekanti}
\affiliation{Department of Physics, University of Toronto, Toronto, Ontario M5S 1A7, Canada}
\affiliation{Canadian Institute for Advanced Research, Toronto, Ontario, M5G 1Z8, Canada}

\date{\today}

\begin{abstract}
Motivated by recent experiments on Bi$_3$Mn$_4$O$_{12}$(NO$_3$), we study a frustrated $J_1$-$J_2$
Heisenberg model on the two dimensional (2D) honeycomb lattice.
The classical $J_1$-$J_2$ Heisenberg model on the 2D honeycomb 
lattice exhibits N\'eel order for $J_2\!\!<\!\! J_1/6$. For $J_2 \!\! > \!\!J_1/6$, it has a family of degenerate
incommensurate spin spiral ground states where the
spiral wave vector can point in any direction. Spin wave
fluctuations at leading order lift this accidental degeneracy in favor 
of specific wave vectors, leading to spiral order by disorder. 
For spin $S=1/2$, quantum fluctuations are, however, likely to be strong
enough to melt the spiral order parameter over a wide range of $J_2/J_1$.
Over a part of this range, we argue that the resulting state is
a valence bond solid (VBS) with 
staggered dimer order - this VBS is a lattice nematic which breaks
lattice rotational symmetry.
Our arguments are supported by comparing the spin wave energy
with the energy of the VBS obtained using a bond operator formalism.
Turning to the effect of 
thermal fluctuations on the spiral ordered state, any nonzero temperature
destroys the magnetic order, but the discrete rotational symmetry of the 
lattice remains broken
resulting in a thermal analogue of the nematic VBS.
We present arguments, supported by classical Monte Carlo 
simulations, that this nematic transforms
into the high temperature paramagnet via a thermal phase transition which is in the universality class of the classical 3-state
Potts (clock) model in 2D.
We discuss the
relevance of our results for honeycomb magnets, such as
Bi$_3$M$_4$O$_{12}$(NO$_3$) (with M=Mn,V,Cr),
and bilayer triangular lattice magnets.
\end{abstract}

\maketitle

\section{Introduction}

Frustrated quantum magnets support a variety of remarkable ground states which
emerge as a result of quantum fluctuations within a large set of classically degenerate configurations.\cite{reviews} Such ground 
states include valence bond solids, magnetic analogues of 
supersolids, and quantum spin liquids with various kinds of topological order. 
While N\'eel order is common in bipartite lattices, the presence of further
neighbor interactions can frustrate this order and lead to interesting
quantum ground states. This has been extensively studied on the square 
lattice \cite{henley1989,chandra1990,sachdev1990,capriotti,weber2003,sirker2006,richter2008}, 
and, for $S=1/2$, there is an indication of a non-magnetic ground state
(for $0.45 \lesssim J_2/J_1 \lesssim 0.6$)
sandwiched between two collinear magnetically ordered ground states.
In this paper, we study the $J_1$-$J_2$ Heisenberg model on the honeycomb 
lattice as the simplest model Hamiltonian which incorporates frustration effects in this lattice geometry. The
Hamiltonian for this model is
\begin{equation}
H = J_1 \sum_{\langle ij\rangle} \bS_i \cdot \bS_j + J_2 \sum_{\langle\langle ij \rangle\rangle} \bS_i \cdot \bS_j,
\label{J1J2}
\end{equation}
where $\langle ij \rangle$ denotes nearest neighbor pairs of sites,  $\langle\langle ij \rangle\rangle$ denotes 
next neighbor pairs of sites, and we set $J_1,J_2\!>\!0$.

A summary of the results contained in this paper is as follows. 
We find that the classical ($S\!=\!\infty$) model has a N\'eel ordered 
ground state for $J_2\!<\!J_1/6$. For $J_2\!>\!J_1/6$, this gives way to a one parameter family of classically
degenerate coplanar spin spiral ground states. At ${\cal O}(1/S)$, quantum
fluctuations within this classical manifold pick specific spiral wavevectors, leading to spiral order by disorder.
For spin $S=1/2$, quantum fluctuations at $T=0$ are likely to be
strong enough to wipe out the spiral order parameter over a wide range of $J_2/J_1$.
Over a significant part of this range of $J_2/J_1$,  we argue that the spiral order for
spin $S=1/2$ melts into a valence bond solid with 
staggered dimer order - this state has a spin gap and 
preserves translational symmetry but breaks
lattice rotational symmetry leading to a `lattice nematic'.
Turning to physics at nonzero temperature, spin-spin correlations decay exponentially 
at any temperature, but we show that the nematic order survives -
this nematic transforms
into the symmetric high temperature 
paramagnet via a thermal phase transition which is in the universality class of the classical 
3-state Potts (clock) model in two dimensions. Some of
the results on the classical degeneracy and spin wave
fluctuations have been discussed earlier,\cite{rastelli1979,fouet2001}
but are included for completeness and clarity.
We also discuss the connection of our work to
previous work on this model 
\cite{oitmaa1978,rastelli1979,morita1986,young1989,oitmaa1991,einarsson1991,
mattsson1994,fouet2001,takano2006}
and related models.\cite{henley1989,chandra1990,weber2003}

Before we get into the detailed analysis of the above model,
we briefly discuss possible materials which might realize the physics
discussed in this paper.
Bi$_3$Mn$_4$O$_{12}$(NO$_3$) appears to be an example of a honeycomb 
lattice quantum
magnet.\cite{BiMnO} Since Mn forms MnO$_6$ octahedral units, and there is strong Hund's coupling, the
Mn$^{4+}$ ions behave as $S=3/2$ spins. Despite the bipartite nature of the lattice, and a large
antiferromagnetic Curie-Weiss constant $\Theta_{CW}\approx -257K$, this system shows no
magnetic order down to $T=0.4K$.\cite{BiMnO} It has been suggested that this
arises from frustration due to further neighbor interactions.\cite{BiMnO} 
Neutron scattering studies would
be valuable to clarify whether such next neighbor couplings are present and whether
they place this system in the regime of fluctuating spiral order, leading to interesting spin liquid behavior 
over a wide range of temperatures, or if there is a nematic transition with its specific heat signature
being obscured by background lattice contributions.
Variants of this system, where Mn$^{4+}$ is replaced by V$^{4+}$ (with $S=1/2$) or by Cr$^{4+}$ (with $S=1$)
would also be interesting to study, with the V$^{4+}$ material being a possible candidate for observing the
dimer solid discussed in this paper.

Among other honeycomb materials,
InCu$_{2/3}$V$_{1/3}$O$_3$
has 
spin-1/2 Cu$^{2+}$ ions nominally forming
a honeycomb lattice \cite{ICVO} 
with the nonmagnetic V$^{5+}$ ions lying at the center of the honeycomb hexagons. However,
this system appears to have strong structural disorder since V$^{5+}$ and Cu$^{2+}$ do not order perfectly in this 
fashion.
Ingredients for other such honeycomb spin systems could be, for instance,
Cu$^{2+}$ ions within CuO$_5$ units arranged on the honeycomb lattice.
In such a trigonal bipyramidal crystal field
environment of the oxygens, the copper ion would then have a single hole, which is located in the $d_{3z^2-r^2}$ orbital.
If the resulting $S=1/2$ moments have significant next neighbor interactions, they might also be candidates
to explore the physics discussed here.

Our study is also relevant to bilayer triangular antiferromagnets where the triangular
layers have an AB stacking,
with antiferromagnetic exchange couplings present between neighboring sites within each layer ($J_2$) 
as well as between neighboring sites
across the two layers ($J_1$). In
this case, each layer acts as one sublattice of the honeycomb antiferromagnet. Such a structure occurs
in LuCuGaO$_4$ which has copper/gallium ions arranged randomly
in a bilayer triangular lattice leading to a strongly disordered spin 
liquid.\cite{LCGO}
A variant such as HfCu$_2$O$_4$, if it could be synthesized in
this structure, might be an interesting material to study.
 
This paper is organized as follows. We begin, in Section II, with a study of the classical model and its many
degenerate spiral ground states and follow it up with an analysis of spin wave fluctuations and how it
selects certain spiral ordered ground states from this manifold. We argue
that spin wave fluctuations are likely
to melt the order for $S\!=\!1/2$ over a wide range of $J_2/J_1$. Section III contains a bond operator approach
to the energetics of the nematic (staggered) dimer solid on the honeycomb lattice. Section IV describes the effect of
thermal fluctuations on such a nematic state using Landau theory as well as by direct Monte Carlo
simulations of the classical $J_1$-$J_2$ Heisenberg model. Section V contains a discussion of 
earlier work on this model and related models on other lattices which share some of the features of the
honeycomb model we have studied.

\section{Spiral order from quantum disorder}

\subsection{Degeneracy of coplanar classical ground states}

To calculate the classical ground state energy we begin by assuming coplanar spiral order on the lattice and
parameterizing the spins on the two sublattices as
\bea
\bS_1(\br) &=& S \left[ \cos(\bQ \cdot \br) \hat{z} + \sin(\bQ  \cdot \br) \hat{x} \right ] \\
\bS_2(\br) &=& - S \left[ \cos(\bQ \cdot \br + \phi) \hat{z} + \sin(\bQ \cdot \br + \phi) \hat{x} \right]
\eea
where $\bQ$ is the spiral wavevector, $\br$ denotes sites on the triangular lattice basis, and $\phi+\pi$
is the angle between spins on the different sublattices at the same site $\br$. This notation is chosen
so that the N\'eel state corresponds to $\bQ=(0,0)$ and $\phi=0$, with spins aligned along $\pm \hat{z}$.

\begin{figure}[bt]
\includegraphics[width=2.8in]{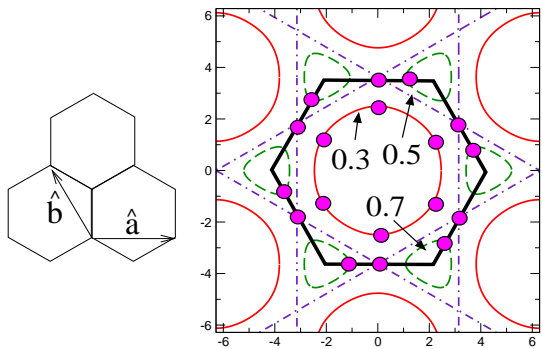}
\caption{(Color online)
{\bf Left panel}: Real space basis vectors for the honeycomb lattice. {\bf Right panel}:
Momentum space picture depicting the 
manifold of classically degenerate spiral wavevectors for $J_2/J_1\!=\!0.3$ (red, thin solid), $J_2/J_1\!=\!0.5$ 
(purple,
dash-dotted), and $J_2/J_1\!=\!0.7$ (green, dashed). Also indicated by purple (solid) dots are the six distinct spiral
wavevectors lying on this manifold which are favored by quantum fluctuations. Black (thick solid) hexagon indicates
the first Brillouin zone of the lattice.}
\label{fig:spiralQ}
\end{figure}

The classical ground state energy per spin is given by
\bea
\frac{E_{\rm cl}}{N}
\!\!\!&\!\!= \!\!\!&\!\!-\frac{J_1 S^2}{2} 
\!\!\left[ \cos \phi \!+\!\cos(\phi\!-\!Q_b)\!+\!\cos(\phi\!-\!Q_a\!-\!Q_b) \right]  \nn \\
&+& J_2 S^2 \left[ \cos Q_a+\cos Q_b + \cos(Q_a+Q_b) 
\right],
\eea
where $\hat{a}=\hat{x}$, and $\hat{b}=-\hat{x}/2+\hat{y}\sqrt{3}/2$, are
unit vectors depicted in Fig.~(\ref{fig:spiralQ}).
Minimizing this classical energy, we find that the minimum energy
solution for $J_2/J_1<1/6$ corresponds to $\bQ^*=(0,0),\phi^*=0$,
so that the N\'eel state is stable for this range of frustration.

For $J_2/J_1>1/6$, the minimum energy solutions correspond to
$\bQ^*$ satisfying the relation
\be
\cos Q^*_a + \cos Q^*_b + \cos (Q^*_a + Q^*_b) = \frac{1}{2} \left[(\frac{J_1}{2 J_2})^2-3\right],
\label{eq:spiralQ}
\ee
while $\phi^*$ is determined completely by
\bea
\sin\phi^* &=& 2 \frac{J_2}{J_1} (\sin Q^*_b + \sin(Q^*_a + Q^*_b)), 
\label{eq:phi1}\\
\cos\phi^* &=& 2 \frac{J_2}{J_1} (1+\cos Q^*_b + \cos (Q^*_a+Q^*_b)).
\label{eq:phi2}
\eea
It is clear that the spiral wavevector is not uniquely fixed by the above relations, as has been noted
earlier.\cite{rastelli1979,fouet2001}
As shown in Fig.\ref{fig:spiralQ},
the set of classically degenerate solutions to
Eq.(\ref{eq:spiralQ}) (which we label $\bQ^*$) form
a closed contour \cite{rastelli1979,fouet2001} around
$\bQ\!\!=\!\!(0,0)$ for $1/6\!<\!J_2/J_1\!<\!1/2$. For $J_2/J_1 \!>\! 1/2$, it forms closed contours around
$(Q_a,Q_b)\equiv \pm(2\pi/3,2\pi/3)$. This regime does not appear to have been investigated in
earlier work. In the limit $J_2/J_1 \!\!\to\!\!\infty$, where
the two triangular sublattices of the honeycomb lattice approximately decouple, 
$\bQ^* \!\to\! \pm(2\pi/3,2\pi/3)$ which is the
ordering wavevector of the $120^\circ$ state on the triangular lattice.
We focus next, therefore, on how quantum or thermal fluctuations select specific
spin spirals from the manifold of classically degenerate coplanar spirals discussed above.

\subsection{Spin wave fluctuations}

We calculate leading quantum 
corrections using Holstein-Primakoff (HP) spin wave theory. We begin
by defining new spin operators $\tilde{\bS}$ via
\begin{equation}
\begin{pmatrix} \tilde{S}^x_{\ell} (\br) \\ \tilde{S}^y_{\ell}(\br) \\ \tilde{S}^z_{\ell} (\br)\end{pmatrix}  = 
\begin{pmatrix} \cos \theta_{\ell}(\br) & 0 & -\sin \theta_{\ell}(\br) \\ 0 & 1 & 0 \\ \sin \theta_{\ell}(\br) & 0 & 
\cos\theta_{\ell}(\br) \end{pmatrix}  \begin{pmatrix} S^x_{\ell}(\br) \\ S^y_{\ell}(\br) \\ S^z_{\ell}(\br) \end{pmatrix}
\label{quantizable_basis}
\end{equation}
where $\ell=1,2$ labels the sublattice, $\theta_1(\br)=\bQ\cdot \br$, and $\theta_2(\br)=\bQ\cdot \br + 
\phi$.
Reexpressing the Hamiltonian in terms of these new spin operators and rewriting these spin operators in
terms of HP bosons, we arrive at the following Hamiltonian which includes 
the leading spin wave correction to
the classical ground state energy,
\begin{equation}
H_{\rm qu} = E_{\rm cl} + 2 S \sum_{\bk>0} \left[
{ \vec{b}_\bk^{\dg} M^\pdg_\bk \vec{b}^\pdg_\bk }
- 2 A_\bk \right].
\end{equation}
Here
$\vec{b}^\dagger = \begin{pmatrix} b_{1}^\dg(\bk)  \ b_2^\dg(\bk) \ b_1^{\pdg}(-\bk) \ b_2^{\pdg}(-\bk) \end{pmatrix}$,
$\sum_{\bk>0}$ 
indicates that the sum runs over half the first Brillouin zone (so that $\bk$ and $-\bk$ are not
both included), and the
Hamiltonian matrix $M_\bk$ takes the form
\be
M_\bk = \begin{pmatrix} A_\bk & B_\bk & C_\bk & D_\bk  \\ B^*_\bk & A_\bk & D^*_\bk & C_\bk  \\ 
C_\bk & D_\bk & A_\bk & B_\bk \\
D^*_\bk & C_\bk & B^*_\bk & A_\bk \end{pmatrix}
\label{eq:Mk},
\ee
with explicit expressions for $A_\bk$-$D_\bk$ given in Appendix A.
Diagonalizing this problem via a generalized Bogoliubov transformation, we obtain the spin wave corrected 
ground energy as
\begin{equation}
E_{\rm qu} = E_{\rm cl} + 2 S \sum_{\bk>0} { \left[\lambda_-(\bk) + \lambda_+(\bk) -  2 A_\bk\right] }
\label{sw-spiral-energy}
\end{equation}
The eigenvalues $\lambda_\pm(\bk)$ are given by
\be
\lambda_{\pm}(\bk) = \sqrt{\alpha_\bk \pm \beta_\bk}
\ee
where
\bea
\alpha_\bk \!\!&=&\! A^2_\bk - C^2_\bk + |B_\bk|^2 - |D_\bk|^2, \\
\beta_\bk\!\!&\!\!=\!\!&\!\!\sqrt{4 |A_\bk B_\bk \!-\!C_\bk D_\bk|^2 \!+\! (D_\bk B^*_\bk \!-\! B_\bk D^*_\bk)^2}.
\eea

For $J_2=0$, it is known from quantum Monte Carlo simulations that this model has long
range N\'eel order.\cite{young1989}
We have checked that the N\'eel state energy for $S=1/2$ is, for $J_2=0$, 
in good agreement with  recent quantum Monte
Carlo simulations in the valence bond basis.
\cite{baskaran2009}

The quantum correction to the classical ground state energy is responsible for selecting a unique quantum ground
state from the manifold of classically degenerate ground states. Minimizing this energy correction over the classical 
ground state manifold, $\bQ^*$,
we find the following results for the
spiral wavevector $\bQ^{**}$, which is selected by quantum fluctuations,
with the resulting $\phi^{**}$ being determined by
Eqns.(\ref{eq:phi1},\ref{eq:phi2}).

\begin{figure}[tb]
\includegraphics[width=2.5in]{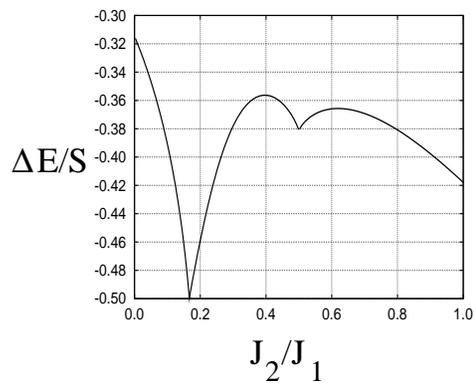}
\caption{Plot of the spin wave correction to the energy per site $\Delta E\! =\! (E_{\rm qu}\!-\!E_{\rm cl})/N$
(in units of $J_1$) as a function of $J_2/J_1$.}
\label{fig:flucn}
\end{figure}

\underline{For $1/6 < J_2/J_1 < 1/2$:} The ground state is a spiral state $S_1$, with 
\bea
Q^{**}_b &=& \cos^{-1}(\frac{J^2_1}{16 J^2_2}-\frac{5}{4})\label{eq:s1a}\\
Q^{**}_a &=& 0 \label{eq:s1b}
\eea
While the above relations specify a single spiral state, there are a total of six 
symmetry related spirals, the other five being
obtained by $2\pi/6$ rotations of the above $\bQ^{**}$.

\underline{For $1/2 < J_2/J_1 < \infty$:} The ground state is a different spiral state $S_2$, with
\bea
Q^{**}_b &=& \pi-\cos^{-1}(\frac{J_1}{4 J_2}+\frac{1}{2}) \\
Q^{**}_a &=& 2  \cos^{-1}(\frac{J_1}{4 J_2}+\frac{1}{2})
\eea
There are six symmetry related $S_2$ spirals, the other five being
obtained by $2\pi/6$ rotations of the above $\bQ^{**}$.
 The spin wave correction to the ground state energy is
shown in Fig.~(\ref{fig:flucn}).

\subsection{Spiral order parameter `melting'}

Spin wave fluctuations will tend to renormalize the spiral order, and may render the spiral states 
unstable. We have checked that the leading spin wave
correction to the spiral order parameter, given by
\bea
\Delta m \!=\! 
S \! -\! \frac{1}{N/2}\sum_{\br }\la {\tilde{S}_{\ell}^z (\br)} \ra \!=\! \frac{1}{N/2}\sum_{\bk} \la b_{\ell}^\dg (\bk) b_{\ell} (\bk)  \ra,
\eea
diverges as $\log(N)$ since the spin wave energy vanishes on the entire classical manifold of degenerate spiral
wavevectors. This suggests that the spiral order will disappear for any value of $S$. However,
while such line zeroes of the dispersion (`Bose surfaces') are mandated by conservation laws in 
the compressible phase of certain
ring-exchange models,\cite{ringexch2002}
it is not generic in this model, and the spin wave energy is expected to only have gapless
points in momentum space corresponding to those wavevectors which are selected by quantum fluctuations.
We expect spin wave interactions, not included at this stage, to gap out all other wavevectors and stabilize the
spiral order for large enough $S$. This requires a higher order spin wave calculation
(in powers of $1/S$) which is beyond the scope of this paper.
At this stage, we restrict ourselves to noting that an exact diagonalization study 
\cite{fouet2001} of the
spin-1/2 model did not find any evidence of a tendency
towards magnetic ordering over a wide range of $J_2/J_1$ where the classical analysis predicts spiral order.
Based on this, we expect that while spiral order might be stabilized at large $S$ from spin wave interactions,
this order is likely to
`melt' for small spin values, leading to other competing states. 

\section{Fluctuation induced `lattice nematic' order}

We have seen that quantum fluctuations in a spin wave expansion will tend to strongly suppress and, for small
spin values, perhaps disrupt the spiral order. In accordance with the Mermin-Wagner theorem, thermal fluctuations are similarly expected to melt the spiral order for any nonzero temperature. While such quantum and thermal fluctuations
may restore spin rotational symmetry with exponentially decaying spin correlations, 
there could be persisting broken symmetries in bilinears of the spin operator
(which are obtained, for instance, by taking dot products or cross products of the single spin operators). 
We begin by listing such bilinears in order to see which of them could possibly survive the effect of fluctuations 
that destroy magnetic long range order.

In the ordered spiral state, ignoring spin wave
corrections to the correlation functions, we find,
\bea
\!\!\!\!\bS_1(\br)\!\!\times\!\!\bS_1(\br\!+\!\bR) \!&\!\!=\!\!&\! \bS_2(\br)\!\!\times\!\!\bS_2(\br\!+\!\bR) \!\!=\!
S^2\! \sin(\bQ\!\cdot\!\bR)  \hat{y} \\
\!\!\!\!\bS_1(\br)\!\!\times\!\!\bS_2(\br\!\pm\!\bR) \!&\!\!=\!\!& - S^2 \sin(\bQ\cdot\bR \!\pm\! \phi) \hat{y}.
\eea
Such bilinears therefore preserve lattice translational symmetry but break the rotational invariance
of the lattice. Since solutions $(\bQ,\phi)$ and $(-\bQ,-\phi)$ are related by a global spin rotation, and
spin correlations are short-ranged at nonzero temperature,
such `vector chiralities' are also expected to have exponentially decaying correlations at
nonzero temperature. By contrast,
spin correlations such as
\bea
\bS_1(\br)\!\cdot\!\bS_1(\br\!+\!\bR) \!&\!\!=\!\!&\! \bS_2(\br)\!\cdot\!\bS_2(\br\!+\!\bR) \!\!=\!
S^2 \!\cos(\bQ\!\cdot\!\bR) \\
\bS_1(\br)\!\cdot\!\bS_2(\br\!\pm\!\bR) \!&\!\!=\!\!& - S^2 \cos(\bQ\!\cdot\!\bR \!\pm\! \phi)
\eea
are
invariant under global spin rotations. These correlations 
are clearly invariant under lattice translations, but they
break lattice rotational symmetry. Such a discrete broken symmetry 
may survive even after fluctuations render the spiral state unstable.

Let us focus on nearest neighbor
bonds and write out the above spin correlations which are simply
proportional to the bond 
energies. We find, for the three bonds around a site on sublattice-1,
\bea
\bS_1(\br)\!\cdot\!\bS_2(\br) &=&  -S^2 \cos\phi\\
\bS_1(\br)\!\cdot\!\bS_2(\br\!-\!\hat{b}) &=& -S^2 \cos(Q_b-\phi)\\
\bS_1(\br)\!\cdot\!\bS_2(\br\!-\!\hat{a}\!-\!\hat{b}) &=& -S^2 \cos(Q_a+Q_b-\phi)
\eea
Computing these bond energies in the $S_1$ spiral ground states selected by quantum fluctuations,
we find that two of the three bond energies are equal while the
third takes on a different value, so that
the $C_3$ rotational symmetry about a lattice site is broken in the $S_1$ state. This is the three-fold symmetry 
that we expect may still be broken even if spin rotational symmetry is restored by quantum or thermal
fluctuations. The $S_2$ state also breaks the three-fold lattice rotational symmetry, as 
seen from the corresponding spiral ordering wavevectors. 
Fluctuations about the spiral states could thus lead to a `lattice nematic' state, which is invariant under 
lattice translations but not lattice rotations. Below, we discuss a quantum nematic valence bond solid 
(VBS) state as a candidate ground state for $S=1/2$, as well a classical nematic state induced by thermal 
fluctuations for any spin value.

\subsection{Nematic valence bond solid}

\begin{figure}[tb]
\includegraphics[width=2in]{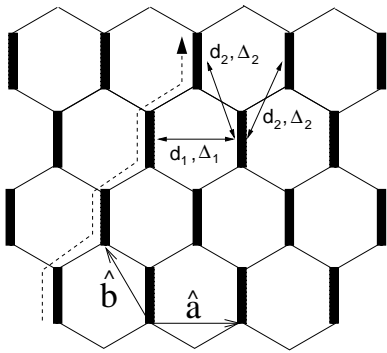}
\caption{Sketch of the valence bond solid state with
staggered dimer order which breaks the honeycomb lattice 
rotational symmetry but preserves spin rotational and lattice
translational symmetries. $\hat{a},\hat{b}$ denote basis (unit) vectors of the triangular
lattice formed by the dimer bonds. $d_{1,2}$ and $\Delta_{1,2}$ indicate
Hartree-Fock-Bogoliubov mean field parameters in the bond operator
mean field theory (see text for details). Dashed line indicates the set of
spins which might be viewed as forming one dimensional dimerized chains
which are coupled in the transverse direction.}
\label{fig:dimer}
\end{figure}

Motivated by the above discussion, we consider the simplest candidate for a lattice nematic ground state for
$S=1/2$ spins, which
corresponds to forming a nematic Valence Bond Solid (VBS) which consists of singlet dimers on the honeycomb lattice 
as
shown in Fig.\ref{fig:dimer}. Such a state has been proposed earlier over a small window of $J_2/J_1$
on the 
basis of a small system exact diagonalization study.
\cite{fouet2001}
An amusing way to
view this VBS state, as shown in Fig.~(\ref{fig:dimer}), is to think of it as
arising from coupling together frustrated spin $S=1/2$ $J_1$-$J_2$ chains.
If we imagine the interchain couplings being tuned to zero, this would 
lead to decoupled
Majumdar-Ghosh chains,\cite{MG}
which are known to possess dimer order with a spin
gap; in particular, the dimerized state is the exact ground state of the
single chain at
$J_2=0.5 J_1$. 
The honeycomb lattice VBS might then be thought of as arising from
the decoupled chain limit upon incorporating interchain couplings while
leaving the singlet gap intact. The choice of which direction these chains
run along is completely arbitrary in the honeycomb limit, so that there are
three degenerate ground states that break the $C_3$ lattice rotational
symmetry.
We note that Heisenberg
models with multispin interactions have been proposed for which this 
VBS state is the exact ground state.\cite{brijesh2009}

In a state where such singlets 
are forced to occur on the indicated bonds, the only excitations correspond to breaking 
these singlets to form triplet excitations which can then form a `triplon' 
band. 
To analyze the energetics and stability of such a state, we therefore use the bond operator formalism proposed in 
Ref.~\onlinecite{sachdev-bhatt1990}.
The details of the calculation are presented in Appendix B.

\begin{figure}[tb]
\includegraphics[width=2.5in]{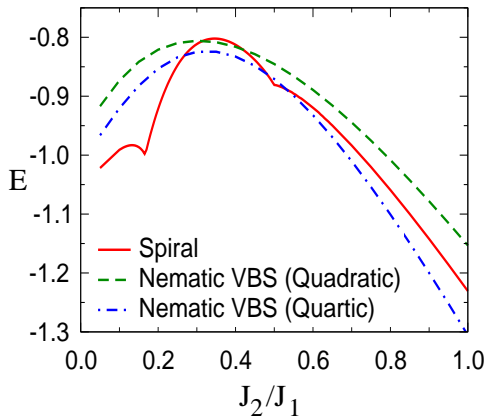}
\caption{ (Color online) Ground state energy (in units of $J_1$)
as a function of $J_2 / J_1$. The red (solid) line is the energy of the 
spiral state including leading order spin wave corrections,
the green (dashed) line is the nematic VBS energy up to quadratic order in triplon operators, and the 
blue (dash-dotted) line indicates nematic VBS energy up to quartic order in triplon operators.}
\label{fig:GSegy}
\end{figure}

Instead of working in the na\"{i}ve basis of $S=1/2$ operators on every site, we switch to a basis of singlet 
and triplet bosonic operators defined as
\bea
s^\dg \vert 0 \ra &=& \frac{1}{\sqrt{2}} (\vert \upa \dna \ra - \vert \dna\upa \ra ) \\
t_{x}^\dg \vert 0\ra &=& \frac{-1}{ \sqrt{2} } (\vert \upa \upa \ra -\vert \dna \dna \ra)  \\
t_{y}^\dg \vert 0\ra &=& \frac{i}{ \sqrt{2} } (\vert \upa \upa \ra +\vert \dna \dna \ra)  \\
t_{z}^\dg \vert 0\ra &=& \frac{1}{ \sqrt{2} } (\vert \upa \dna \ra +\vert \dna \upa \ra)  
\eea
on the dark (dimer) bonds in Fig.\ref{fig:dimer}, together with a local constraint
\be
s_{\br}^\dg s_{\br} + \sum_{\alpha=x,y,z} t_{\br,\alpha} ^\dg t_{\br,\alpha} = 1
\label{constraint}
\ee
Here $\alpha$ is summed over $x,y,z$, $t_\alpha$ being the three triplon operators on the bond at $\br$. 
(Repeated Greek indices henceforth denote summation over $x,y,z$.)
To simplify the calculation, we satisfy this constraint on average, rather than locally. We assume the singlet 
operators to be condensed, allowing us to replace the operator $s_i$ with a number $\sbar$. 
The excitations are the triplet operators, and the terms of the Hamiltonian may now be organized in order of 
the number of triplet operators. 
The Hamiltonian in momentum space, keeping only terms up to quadratic order, is
\bea    
\nn H_{BO}^ {[2]} \!\!&\!=\!&\!\! -\frac{3N}{4}J_1 \sbar^2 - N\mu \sbar^2 + N\mu - 3\sum_{\bk > 0} G_\bk \\
\! &\!+\!&\!\! \sum_{\bk>0}\!\left[\begin{array}{cc}
\!t_{\gamma} ^\dg (\bk) & \!t_{\gamma}(-\bk)\end{array}\right]
\left[\begin{array}{cc}
G_\bk & F_\bk \\
F_\bk^* & G_\bk
\end{array}\right] 
\left[\begin{array}{c}
t_{\gamma} (\bk) \\
t_{\gamma} ^\dg (-\bk)
\end{array}\right]
\label{tripHqd}
\eea      
where $\mu$ is a chemical potential that has been introduced to satisfy the constraint in Eq.~(\ref{constraint}).
The matrix entries are given by 
\bea       
\nn G_\bk &=& \frac{J_1}{4} -\mu -\frac{\sbar^2}{4} J_1 (\epsilon_\bk + \epsilon_{-\bk}) + \frac{\sbar^2}{4} J_2 (\eta_{\bk} + \eta_{-\bk}) \\       
\nn F_\bk &=& -\frac{\sbar^2}{4} J_1 (\epsilon_\bk + \epsilon_{-\bk}) + J_2 \frac{\sbar^2}{4} (\eta_{\bk} + \eta_{-\bk})       
\eea       
where
\bea
\nn \epsilon_\bk &=& e^{-ik_b} + e^{-i(k_b+k_a)} \\
\nn \eta_\bk &=& 2 [\cos(k_a) +  \cos(k_b) +  \cos(k_a + k_b)]
\eea
Diagonalizing the Hamiltonian by a bosonic Bogoliubov transformation gives the dispersion of the eigenmodes to be
\be
E_k = \sqrt{G_k^2 - \vert F_k \vert ^2}
\ee
The energy of this state is plotted as the green dashed line in Fig~(\ref{fig:GSegy}).

While this quadratic theory gives a consistent picture of our lattice nematic state, higher order terms may lower its
energy significantly. 
We proceed to take these into account by means of a self consistent 
Hartree-Fock approach. This approach
has been compared recently, for a star lattice Heisenberg antiferromagnet, with Gutzwiller projected variational wavefunctions 
\cite{bjyang2010} and exact diagonalization studies \cite{richter2004}
and shown to provide a good description of the energetics of valence bond
solid states on the star lattice.\cite{bjyang2010}
We begin by noting that the terms of cubic order in the triplet operators do not contribute, since we work with the 
assumption that the triplon operators themselves are not condensed. The quartic part of the Hamiltonian is given by
\bea
\nn H_{BO} ^{[4]} = \frac{-1}{4N} 
\sum_{\bk,\bk',\bq} 
\epsilon_{\mu\beta\gamma}\epsilon_{\mu\nu\delta}  
(J_1 \epsilon_{\bk-\bk'} + J_2 \eta_{\bk-\bk'}) \\
t_{\beta}^\dg (\bk+\bq) t_{\gamma} (\bk'+\bq) 
t_{\nu}^\dg (\bk') t_{\delta} (\bk)
\label{tripHqr}
\eea
where $\epsilon_{\mu\beta\gamma}$ is the permutation symbol. 
Guided by the symmetry of the nematic phase, we postulate the following real-space order 
parameters:
\bea
d_{1} &=& \frac{1}{3} \la t_{\br,\gamma}^\dg t_{\br + {\bf \delta}_1, \gamma} \ra \\
d_{2} &=& \frac{1}{3} \la t_{\br,\gamma}^\dg t_{\br + {\bf \delta}_2,\gamma} \ra \\
\Delta_{1} &=&  \frac{1}{3} \la t_{\br,\gamma} t_{\br + {\bf \delta}_1,\gamma} \ra \\
\Delta_{2} &=& \frac{1}{3} \la t_{\br,\gamma} t_{\br + {\bf \delta}_2,\gamma} \ra
\label{qrOPs}
\eea
where
$\delta_1=\pm \hat{a}$, and $\delta_2 =\pm \hat{b},\pm(\hat{a}+\hat{b})$.
These are defined on bonds as shown in Fig.~(\ref{fig:dimer}). The above are the only operators that couple to 
$\sbar^2$ at quadratic level.

We calculate these order parameters 
self-consistently, and thereby obtain the energy of the nematic VBS, having accounted for quartic terms. This is plotted in 
Fig.~(\ref{fig:GSegy}) as the blue (dot-dashed) line. At quadratic level, the nematic state is energetically favourable over the spiral over a small window
near $J_2\sim 0.35 J_1$.  Although we have not considered interactions between spin wave modes, it is
nevertheless, it is encouraging that the quartic level energy in the bond operator formalism 
is lower than the spin wave energy for
$J_2 \gtrsim 0.25 J_1$, 
except for a small window around $J_2 = 0.5 J_1$. Since the spiral order is anyway likely to be suppressed by
fluctuations, our results are quite suggestive of such nematic VBS order being present over
a wide window of frustration. At large $J_2/J_1$, we expect competing states might emerge which
are descendants of spin liquid states on the triangular lattice \cite{wang2006} - this needs further investigation.

We note that this bond operator formalism does
not take into account the fluctuations of the singlets themselves; the
kinetic energy lowering from such resonating singlet valence bonds 
might possibly
favor $\sqrt{3}\times\sqrt{3}$ plaquette dimer order which also breaks
lattice translation symmetry \cite{sachdev1990,moessner2001} - such states can be accessed within a 
Schwinger boson formalism and might be relevant in the vicinity of the
point where the N\'eel order is lost.

\subsection{Thermal fluctuations: Landau theory}

The spiral states $S_1$ and $S_2$, obtained from including spin fluctuations at large S, can only be stable at zero temperature. At any non-zero temperature, since our system is two dimensional, spin rotational symmetry will be immediately restored. As earlier discussed, the simplest ordering would involve nearest-neighbor bilinears of the spin
operators, which may break lattice rotational symmetry. Motivated by earlier work on such `lattice nematics' and quantum dimer models, 
we define a local complex order parameter
\bea
\psi(\br) &=&  \langle \bS_1(\br)\!\cdot\!\bS_2(\br)\rangle + \omega \langle \bS_1(\br)\!\cdot\!\bS_2(\br\!-\!\hat{b}) \rangle\nn\\
&+& \omega^2  \langle \bS_1(\br)\!\cdot\!\bS_2(\br\!-\!\hat{a}\!-\!\hat{b})\rangle
\eea
on sites of sublattice 1, where $\omega=\exp(i 2\pi/3)$. Since two of the bond energies are equal in
the ground state, $\psi(\br) \sim \{1,\omega,\omega^2\}$ in the three ground states, so that
$\psi^3(\br) \sim 1$. This order parameter is invariant under translations even in the spiral state. Under
an anticlockwise $2\pi/3$ rotation about a site on sublattice 1, we have $\psi \to \omega \psi$.
Finally, reflections about axes running along the bonds
can be shown to lead to $\psi \to \psi^*$.
Based on these symmetry considerations, the finite temperature classical lattice nematic
is expected to be described by a Landau free energy functional of the form
\bea
{\cal F} &=& \int \! d^2\br ~\left[m |\psi(\br)|^2 \!+\! u |\psi(\br)|^4 + \lambda |\nabla \psi(\br)|^2 \right. \nonumber\\
&+& \left. w (\psi^3(\br)\!+\!\psi^{* 3}(\br))
\right]
\label{landau}
\eea
which is equivalent to a 3-state clock model (or equivalently, the 3-state 
Potts model) if we assume
that amplitude is fixed.
We therefore expect the classical model (as well as possibly the $S=1/2$ model) to exhibit, upon
warming up from $T=0$, a 
finite temperature transition from a lattice nematic
into an ordinary paramagnet, with this transition being in the universality
class of the 3-state Potts model in 2D. (Of course, these arguments do not rule out the possibility of a first-order
transition.)

\subsection{Thermal fluctuations: Monte Carlo study}

We have carried out Monte Carlo simulations of the
classical $J_1$-$J_2$ Heisenberg model
in order to numerically explore the nematic-paramagnet phase transition. 
\begin{figure}[t]
\includegraphics[width=3in]{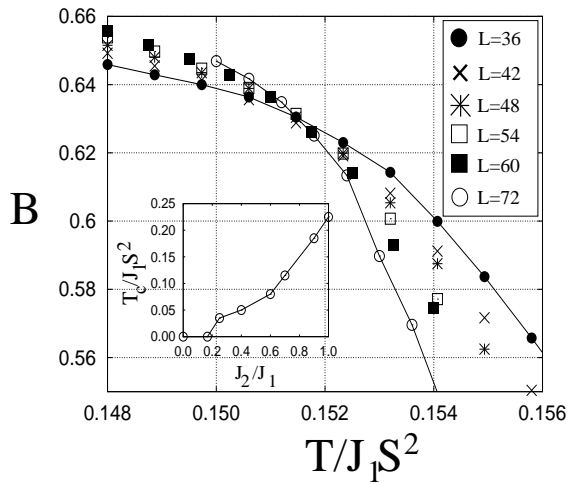}
\caption{
Binder cumulant of the order parameter (from Eq.~\ref{binder}) plotted
as a function of temperature for various system sizes. The crossing point of
the curves at $T/J_1S^2 \approx 0.1515(5)$ indicates a continuous nematic
to paramagnet phase 
transition at $J_2/J_1\!=\!0.8$. Inset shows the nematic transition temperature, $T_c$, as a
function of $J_2/J_1$. Lines are guides to the eye.}
\label{fig:Tc}
\end{figure}
Using a combination of
single-spin Metropolis moves and energy conserving (over-relaxed) moves, we have computed the Binder cumulant 
of the order parameter,
\be
B =1-\frac{\la |\Psi|^4 \ra}{3 \la|\Psi|^2\ra^2},
\label{binder}
\ee
where $\Psi= \sum_{\br} \psi(\br)$ is a complex scalar, and the susceptibility,
\be
\chi = \frac{1}{N T} (\la |\Psi^2| \ra - \la |\Psi| \ra^2),
\label{chi}
\ee
for the classical Heisenberg model for various $J_2/J_1$. Fig.~\ref{fig:Tc}
shows the Binder cumulant as a function
of temperature obtained on various system sizes ($N\!=\!L^2$ with 
$L\!=\!36$-$72$) for $J_2/J_1=0.8$ by
averaging over $10^6$-$10^7$ configurations.
These exhibit a crossing point at $T_c/J_1 S^2 \approx 0.1515(5)$ indicating a continuous thermal phase transition, with 
$B(T_c) \approx 0.63$.

\begin{figure}[tb]
\includegraphics[width=3in]{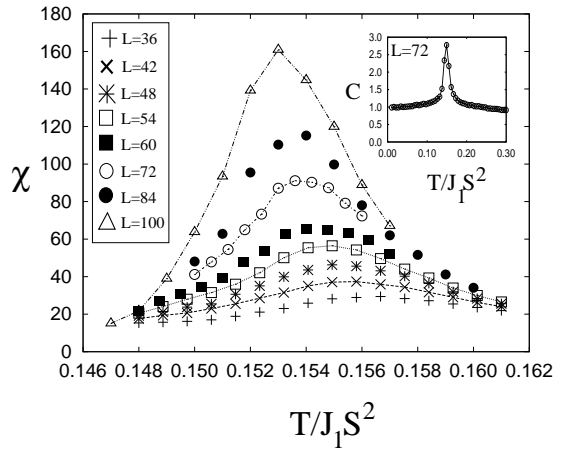}
\caption{Nematic susceptibility (from Eq.~\ref{chi}), in units of $1/J_1$,
as a function of temperature for various 
system sizes at $J_2/J_1=0.8$.  Lines are guides to the eye. 
Inset shows the specific heat peak at the transition for $L=72$.}
\label{fig:chi}
\end{figure}

In addition, as seen from Fig.~\ref{fig:chi},
the peak of the susceptibility (at $J_2/J_1\!=\!0.8$)
increases with system size. Based on the finite size scaling of this peak height, 
$\chi \sim L^{\gamma/\nu}$, we find
$\gamma/\nu \approx 1.68(8)$. The 
order parameter $\la |\Psi| \ra$ at $T_c$ scales  with system size as 
$L^{-\beta/\nu}$,
with $\beta/\nu \approx 0.14(2)$. Finally, the shift in the susceptibility
peak with system size is expected to scale as $\Delta T_\chi \sim L^{-1/\nu}$
from which we find $1/\nu \approx 1.25(9)$. 
These results for the exponents
are reasonably consistent with a 3-state 
Potts model transition for 
which the exact exponents
are known \cite{baxter} 
to be $\beta=1/9$,$\gamma=13/9$, and $\nu=5/6$; these imply
$\gamma/\nu \approx 1.733$, $\beta/\nu \approx 0.133$, and $1/\nu=1.2$.
The critical Binder cumulant $B(T_c)$ also seems consistent with earlier
numerical work on the Potts model.\cite{southern2003}
We have also computed
the specific heat of this model, and, as seen in the inset of Fig.\ref{fig:chi}, 
it shows a clear peak at the transition point located from the Binder
cumulant calculation. For reasons we do not completely understand, the specific heat does not
exhibit clear finite size scaling over the system sizes explored. It is possible that
the thermally fluctuating spin wave 
modes and their interaction with the nematic order parameter
may make it difficult to extract the finite size scaling of the specific heat singularity for system sizes we have
studied.
We are carrying out careful numerical studies of this model
on larger system sizes in order to understand this issue. 

\section{Relation to previous work}

Earlier investigations of the honeycomb lattice model have focused on the spin wave selection of various
spiral states.\cite{rastelli1979,fouet2001} Our results are in line with these studies - it appears that specific
spin spirals are selected at ${\cal O}(1/S)$ in a 
spin wave calculation, but the resulting order is likely
to `melt' for $S=1/2$ over a wide range of $J_2/J_1$. An exact diagonalization study \cite{fouet2001} 
of the spin $S=1/2$
model has suggested that nematic order with breaking of $C_3$ rotational symmetry could appear in the
vicinity of $J_2\!=\!0.4$-$0.5J_1$, and this order has also been guessed from a study of Berry phase effects in
a nonlinear sigma model formulation.\cite{einarsson1991} Our
bond operator calculations lend support to this claim, and also suggest that this nematic dimer order may 
persist over a wide range of $J_2/J_1$.

The idea that isotropic Heisenberg models may have such nematic orders at finite temperature is
well known from early work on the square lattice $J_1$-$J_2$ model.\cite{chandra1990}
For $J_2 < J_1/2$, the classical ($S=\infty$) ground state of this model on the square lattice
is N\'eel ordered. For $J_2 > J_1/2$, by contrast,
there is a large set of classically degenerate ground states in which the two sublattices are individually perfectly
N\'eel ordered with an arbitrary relative angle between the two sublattices. Within this classical manifold,
quantum
fluctuations at ${\cal O}(1/S)$ select collinear ground states 
\cite{henley1989}
with ordering wavevectors $\bQ = (\pi,0)$ or $(0,\pi)$.
At any nonzero temperature, this model exhibits exponentially decaying spin correlations,
consistent with the 
Mermin-Wagner theorem, but the broken lattice rotational symmetry associated with these collinear ground 
states survives at low temperature.
Upon further heating, this `lattice nematic', which breaks the $C_4$ rotational symmetry of the square lattice down
to $C_2$, converts into the high temperature paramagnetic phase via an 
Ising transition.\cite{chandra1990,weber2003}
Despite a large number of numerical studies,\cite{sachdev1990,capriotti,sirker2006,richter2008} 
however, the ground state phase diagram of this square lattice spin-1/2 model 
appears to not to be satisfactorily understood. 

The relation between spiral magnetic states and nematic orders has also been explored in the context of the
$J_1$-$J_3$ model on the square lattice.\cite{capriotti-J1J3} 
In this case, there is a N\'eel to spiral transition for $J_3 > J_1/4$,
which is a Lifshitz transition similar to the case we have studied. Melting this spiral thermally leads to an Ising
nematic similar to the square lattice $J_1$-$J_2$ model. The main differences of our model with this case
are: (i) The spiral wavevector is unique (modulo reflections) in the classical square lattice $J_1$-$J_3$ model
unlike the line degeneracy we encounter on the honeycomb lattice; (ii)
the nematic-paramagnet transition in the square lattice $J_1$-$J_3$ model is an Ising transition;
(iii) unlike on the honeycomb
lattice, there is no simple quantum analogue of the classical nematic in the square lattice model. It may be
more useful to consider possible analogies of the honeycomb model with 
the $J_1$-$J_2$-$J_3$ model on the square lattice which has been studied in recent work.\cite{sindzingre2010}

Some features
of the honeycomb model, such as a highly degenerate set of classical 
spiral states and the resulting spiral selection by fluctuation
effects, bear similarities with studies 
on the diamond lattice $J_1$-$J_2$ model,\cite{balents2007,ybkim2008}
which were motivated by
insulating 
spinel compounds such as MnSc$_2$S$_4$, Co$_3$O$_4$, and CoRh$_2$O$_4$.
We note, in passing, that the issue of spiral order and its connection
to lattice nematicity also arises in the context of itinerant systems. In
particular, such spiral melting has been proposed as one mechanism 
\cite{simons2009} for the
observed nematic transport \cite{mackenzie2007} in
Sr$_3$Ru$_2$O$_7$ at intermediate magnetic fields,
although there are
competing theoretical proposals \cite{nematics}
for the observed nematic order.

\section{Summary}
We have studied the honeycomb lattice $J_1$-$J_2$ Heisenberg model. We have seen that the classical
model supports a one-parameter family of degenerate spin spiral states, of which specific spin spirals are
selected out by quantum fluctuations. 
For general spin values, we expect the spiral order to be strongly
suppressed but robust nematic order to survive. For $S=1/2$, spin fluctuations are likely strong enough 
to `melt' the spiral order leading to a spin gapped nematic dimer solid as indicated from our bond operator 
calculations. We have shown that the classical model, and possibly also the dimer solid, are connected to the high 
temperature paramagnetic phase via a 3-state Potts model transition. 
Neutron scattering experiments would be valuable to test for
fluctuating spiral order at finite temperatures -
in this regime, the equal time structure factor
exhibits peaks on the spiral contours in Fig.~\ref{fig:spiralQ} 
as the system thermally explores the various (nearly) degenerate spirals.
This may allow a determination of the further neighbor couplings which
frustrate N\'eel order. Further work is necessary to determine if interesting
gapless spin liquids emerge as candidate ground states for this
model over some regime of frustration as has been recently proposed for
other frustrated quantum magnets.\cite{motrunich,hermele,marston,
palee,lawler,trivedi} The other interesting possibility is the existence
of gapped spin liquids as have been
recently uncovered in numerical studies in the insulating state of the
honeycomb lattice Hubbard
model.\cite{assaad} The model we have studied appears to be directly
applicable as an effective spin Hamiltonian (with $J_2/J_1 \!\approx \!0.1$) in the insulating phase
of the Hubbard model for moderate values of repulsion.
Finally, our results are relevant to 
honeycomb and bilayer triangular magnets; we hope our work 
stimulates further experiments on
the Bi$_3$M$_4$O$_{12}$(NO$_3$) family of materials, and other 
compounds which might realize this model.

\acknowledgments

We thank M. Azuma, A. Banerjee, S. Bhattacharjee,
K. Damle, T. Dodds, Y.-B. Kim, Y.-J. Kim, C. Lhuillier,
D. Podolsky, T. Senthil, P. Sindzingre, and B.-J. Yang, 
for useful discussions. This research was
supported by NSERC of Canada. AP acknowledges support
from an Ontario ERA and the Sloan Foundation.

\bigskip

\appendix

\section{Matrix Elements for the Holstein-Primakoff Hamiltonian}
The explicit expressions for the matrix elements of the Holstein-Primakoff Hamiltonian in
Eq.(\ref{eq:Mk}) are given by
\bea
A_\bk &=& \frac{J_1}{2} [ \cos \phi \!+\! \cos(\phi \!-\! Q_b) \!+\! \cos(\phi \!-\! Q_a \!-\! Q_b) ] \nn \\
&-& J_2 [ \cos Q_a + \cos Q_b  + \cos(Q_a + Q_b) ] \nn \\ 
&+& \frac{J_2}{2} [(\cos Q_a +1)\cos k_a + (\cos Q_b+1)\cos k_b \nn\\
&+& (\cos(Q_a+ Q_b)+1)\cos(k_a + k_b) ]\\
B_\bk&=& \frac{J_1}{4} [ (\cos \phi -1) + (\cos(\phi - Q_b)-1)e^{-ik_b} \nn \\
&+& (\cos(\phi - Q_a - Q_b)-1)e^{-i(k_a + k_b)} ] \\
C_\bk &\equiv& \Gamma_\bk + \Gamma_\bk^* \\
\Gamma_\bk &=& \frac{J_2}{4} [ (\cos(Q_a)-1)e^{ik_a} + (\cos(Q_b)-1)e^{-ik_b} \nn \\
&+& (\cos(Q_a+Q_b)-1)e^{i(k_a+k_b)}] \\
D_\bk &=& \frac{J_1}{4} [ (\cos \phi+1) + (\cos(\phi - Q_b)+1)e^{-ik_b} \nn \\
&+& (\cos(\phi - Q_a - Q_b)+1)e^{-i(k_a + k_b)} ]
\eea

\section{Triplon Calculation for Lattice Nematic State}
We work with a basis of singlet and triplet operators that are centred on bonds indicated in Fig.~(\ref{fig:dimer}).
In terms of the bond operators defined in the text, the spin operator on any site can be written as 
\be
{S}_{\ell}^\gamma (\br) = \frac{1}{2} f(\ell) (s_{\br}^\dg t_{\br,\gamma} + t_{\br,\gamma}^\dg s_{\br}) - \frac{i}{2} \epsilon_{\gamma\beta\delta} t_{\br,\beta}^\dg t_{\br,\delta}
\ee
Here, the index $\ell$ indicates the sublattice. The factor $f(\ell)$ takes the value $+1$ on sublattice 1 and $-1$ on sublattice 2. $\br$ is summed over sites of the bond-centred triangular lattice. 

We now consider the singlets to have condensed, giving us a nematic state. Between sites that are connected by a bond, we have 
\be
{\bf S}_{1} (\br) . {\bf S}_{2} (\br) = -\frac{3}{4} \sbar^2 + \frac{1}{4}\sum_{\gamma} t_{\br,\gamma}^\dg t_{\br,\gamma}
\ee
For sites that are not connected by a bond, we have 
\bea
\nn {\bf S}_{\ell} (\br).{\bf S}_{\ell'}(\br') = f(\ell) f(\ell') \frac{\sbar^2 }{4} (t_{\br,\gamma}^\dg + t_{\br,\gamma})\times \\
\nn ( t_{\br',\gamma}^\dg + t_{\br',\gamma}) 
- \frac{i\sbar}{4}\epsilon_{\gamma\beta\delta} [ f(\ell) (t_{\br,\gamma}^\dg + t_{\br,\gamma})t_{\br',\beta}^\dg t_{\br',\delta} + \\
\nn t_{\br,\beta}^\dg t_{\br,\delta} f(\ell')(t_{\br',\gamma}^\dg + t_{\br',\gamma}) ] \\
+ \frac{(-1)}{4}\epsilon_{\gamma\beta\delta}\epsilon_{\gamma\nu\eta}t_{\br,\beta}^\dg t_{\br,\delta} t_{\br',\nu}^\dg t_{\br',\eta}
\eea
To enforce the constraint on the bond operators, we rewrite the Hamiltonian as
\be
H - \mu \sum_{\br}\left[ \sbar^2 + t_{\br,\alpha}^\dg t_{\br,\alpha} -1  \right]
\ee
where $H$ is the original spin Hamiltonian in Eq.~(\ref{J1J2}). $\mu$ is now tuned so that the constraint is satisfied on average. Keeping terms to quadratic order in the $t$-operators, 
this Hamiltonian may be rewritten as Eq.~(\ref{tripHqd}), and diagonalized by a Bogoliubov transformation. 
For fixed $J_2/J_1$, we choose the value of $\sbar$ that minimizes energy. 


The term that is quartic in triplon operators is given by Eq.~(\ref{tripHqr}). This can be decoupled in hopping and pairing channels, using the order parameters defined in Eq.~(\ref{qrOPs}). 
This modifies the coefficients of the quadratic Hamiltonian of Eq.~(\ref{tripHqd}) as follows
\bea
G_\bk ^{[4]} &=& G_\bk + \frac{ J_1}{2} d_{2}(\epsilon_\bk + \epsilon_{-\bk}) + { J_2} d_{2}(\epsilon_\bk + \epsilon_{-\bk}) \nn \\
&+& 2{ J_2} d_{1}\cos(k_a)  \\
F_\bk ^{[4]} &=& F_\bk - \frac{ J_1}{2} \Delta_{2}(\epsilon_\bk + \epsilon_{-\bk}) - { J_2} \Delta_{2}(\epsilon_\bk + \epsilon_{-\bk}) \nn \\
&-& 2 J_2 \Delta_{1}\cos(k_a)
\eea
In addition, the Hamiltonian acquires a constant contribution given by
\bea
\delta E^{[4]} = -{3 J_1}N (d_{2}^2 - \vert \Delta_{2} \vert^2) -{ 6 J_2}N(d_{2}^2 - \vert \Delta_{2} \vert^2) \nn \\
-{3 J_2}N (d_{1}^2 - \vert \Delta_{1} \vert^2)
\eea
The Hamiltonian is solved by a Bogoliubov transformation. For fixed $\sbar$, the $d$ and $\Delta$ order parameters are determined self-consistently, while $\mu$ is tuned to make sure the constraint on bond operators is satisfied. $\sbar$ is
chosen to minimize the ground state energy for every $J_2/J_1$.


\begin{thebibliography}{999}
\bibitem{reviews}
L. Balents, Nature {\bf 464}, 199 (2010);
A. P. Ramirez, Nat. Phys. {\bf 4}, 442 (2008);
P. A. Lee, Rep. Prog. Phys. {\bf 71}, 012501 (2008);
R. Moessner and A. P. Ramirez, Physics Today, 24 (2006);
G. Misguich and C. Lhuillier, in ``{\it Frustrated Spin Systems}'',
edited by H. T. Diep (World Scientific, New York, 2005).
\bibitem{henley1989}
C. Henley, Phys. Rev. Lett. {\bf 62}, 2056 (1989).
\bibitem{chandra1990}
P. Chandra, P. Coleman, and A. I. Larkin, Phys. Rev. Lett. {\bf 64}, 88 (1990). 
\bibitem{sachdev1990}
N. Read and Subir Sachdev, Phys. Rev. Lett. 62, 1694 (1989); 
N. Read and S. Sachdev, Phys. Rev. B 42, 4568 (1990).
\bibitem{capriotti}
L. Capriotti and S. Sorella, Phys. Rev. Lett. 84, 3173 (2000);
L. Capriotti, F. Becca, A.Parola, and S. Sorella, \prl {\bf 87}, 097201 (2001);
F. Becca, L. Capriotti, A. Parola, S. Sorella, \prb {\bf 76}, 060401 (2007).
\bibitem{weber2003}
C. Weber, L. Capriotti, G. Misguich, F. Becca, M. Elhajal, and F. Mila,
Phys. Rev. Lett. {\bf 91}, 177202 (2003).
\bibitem{sirker2006}
J. Sirker, Z. Weihong, O.P. Sushkov, and J. Oitmaa, Phys. Rev. B {\bf 73}, 184420 (2006).
\bibitem{richter2008}
R. Darradi, O. Derzhko, R. Zinke, J. Schulenburg, S. E. Krüger, and J. Richter
Phys. Rev. B 78, 214415 (2008).
\bibitem{rastelli1979}
E. Rastelli, A. Tassi, and L. Reatto, Physica B {\bf 97}, 1 (1979).
\bibitem{fouet2001}
J. B. Fouet, P. Sindzingre, and C. Lhuillier,
European Physical Journal B {\bf 20}, 241 (2001).
\bibitem{oitmaa1978}
J. Oitmaa and D. D. Betts, Can. J. Phys. {\bf 56}, 897 (1978)
\bibitem{morita1986}
S. Katsura, T. Ide, and Y. Morita, J. Stat. Phys. {\bf 42}, 381 (1986).
\bibitem{young1989}
J. D. Reger, J. A. Riera, and A. P. Young, J. Phys. Cond. Matt.
{\bf 1}, 1855 (1989).
\bibitem{oitmaa1991}
Z. Weihong, J. Oitmaa, and C. J. Hamer, \prb {\bf 44}, 11689 (1991);
J. Oitmaa, C. J. Hamer, and Z. Weihong, \prb {\bf 45}, 9834 (1992).
\bibitem{einarsson1991}
T. Einarsson and H. Johannesson, \prb{43}, 5867 (1991).
\bibitem{mattsson1994}
A. Mattsson, P. Fr\"ojdh, and T. Einarson, \prb {\bf 49}, 3397 (1994).
\bibitem{takano2006}
K. Takano, \prb {\bf 74}, 140402 (2006).
\bibitem{BiMnO}
O. Smirnova, M. Azuma, N. Kumada, Y. Kusano, M. Matsuda, Y. 
Shimakawa, T. Takei, Y. Yonesaki, and N. Kinomura,
J. Am. Chem. Soc., {\bf 131}, 8313 (2009);
S. Okubo, F. Elmasry, W. Zhang, M.
Fujisawa, T. Sakurai, H. Ohta, M. Azuma, O.
A. Sumirnova, and N. Kumada,
J. Phys.: Conf. Ser. {\bf 200}, 022042 (2010).
\bibitem{ICVO}
A. M\"oller, U. L\"ow, T. Taetz,M. Kriener, G. Andr\'e, F. Damay, O. Heyer,  M. Braden, and J. A. Mydosh,
\prb {\bf 78}, 024420 (2008); M. Yehia, E. Vavilova, A. M\"oller, T. Taetz, U. L\"ow, R. Klingeler, V. Kataev, and 
B. B\"uchner,
Phys. Rev. B {\bf 81}, 060414 (2010).
\bibitem{LCGO}
R. J. Cava, A. P. Ramirez, Q. Huang, and J. J. Krajewski, 
J. Solid State Chem. {\bf 140}, 337 (1998); 
S. Calder, S. R. Giblin, D. R. Parker, P. P. Deen, C. Ritter,
J. R. Stewart, and T. Fennell, arXiv:1002.0975 (unpublished).
\bibitem{baskaran2009}
Z. Nourbakhsh, F. Shahbazi, S. A. Jafari, and G. Baskaran,
J. Phys. Soc. Jpn. {\bf 78}, 054701 (2009).
\bibitem{ringexch2002}
A. Paramekanti, L. Balents, and M. P. A. Fisher, \prb {\bf 66}, 054526 (2002).
\bibitem{MG}
C. K. Majumdar and D. K. Ghosh, J. Math. Phys. {\bf 10}, 1388 (1969);
C. K. Majumdar and D. K. Ghosh, J. Math. Phys. {\bf 10}, 1399 (1969);
F. D. M. Haldane, \prl {\bf 25}, 4925 (1982).
\bibitem{brijesh2009}
R. Kumar, D. Kumar, and B. Kumar, Phys. Rev. B {\bf 80}, 214428 (2009).
\bibitem{sachdev-bhatt1990}
S. Sachdev and R. N. Bhatt, \prb {\bf 41}, 9323 (1990).
\bibitem{richter2004}
J. Richter, J. Schulenberg, A. Honecker, and D. Schmalfuss, \prb{\bf 70}, 174454 (2004).
\bibitem{bjyang2010}
B.-J. Yang, A. Paramekanti, and Y. B. Kim, arXiv:0911.2702 (Phys. Rev.
B, to appear).
\bibitem{wang2006} 
F. Wang and A. Vishwanath, \prb {\bf 74}, 174423 (2006). 
\bibitem{moessner2001}
R. Moessner, S. L. Sondhi, P. Chandra, \prb {\bf 64}, 144416 (2001).
\bibitem{baxter}
R. J. Baxter, {\it Exactly Solved Models in Statistical Mechanics},
(Academic, London, 1982).
\bibitem{southern2003}
Z. F. Wang and B. W. Southern, \prb{\bf 68}, 094419 (2003).
\bibitem{capriotti-J1J3}
L. Capriotti and S. Sachdev, \prl {\bf 93}, 257206 (2004).
\bibitem{sindzingre2010}
P. Sindzingre, N. Shannon, T. Momoi,
{\it J. Phys: Conf.Ser}, {\bf 200},  022058 (2010).
\bibitem{balents2007}
D. Bergman, J. Alicea, E. Gull, S. Trebst, and
L. Balents, Nat. Phys. {\bf 3}, 487 (2007). 
\bibitem{ybkim2008}
J.-S. Bernier, M. J. Lawler, and Y. B. Kim, Phys. Rev. Lett. {\bf 101}, 047201 (2008). 
\bibitem{simons2009}
A. M. Berridge, A. G. Green, S. A. Grigera, and B. D. Simons
Phys. Rev. Lett. {\bf 102}, 136404 (2009); 
A. M. Berridge, S. A. Grigera, B. D. Simons, and A. G. Green, 
Phys. Rev. B {\bf 81}, 054429 (2010).
\bibitem{mackenzie2007}
R. A. Borzi, S. A. Grigera, J. Farrell, R. S. Perry, S. J. S.  Lister, 
S. L. Lee, D. A.
Tennant, Y. Maeno, and A. P. Mackenzie, Science {\bf 315}, 214 (2007).
\bibitem{nematics}
H.-Y. Kee and Y. B. Kim, Phys. Rev. B {\bf 71}, 184402 (2005);
S. Raghu, A. Paramekanti, E.-A. Kim, R. A. Borzi, S. Grigera, A. P. 
Mackenzie, and S. A. Kivelson, Phys. Rev. B {\bf 79}, 214402 (2009);
W.-C. Lee and C. Wu, Phys. Rev. B {\bf 80}, 104438 (2009); 
C. M. Puetter, J. G. Rau, and H.-Y. Kee, Phys. Rev. B {\bf 81}, 081105 (2010);
M. H. Fischer and M. Sigrist, Phys. Rev. B {\bf 81}, 064435 (2010).
\bibitem{motrunich}
O. I. Motrunich, Phys. Rev. B {\bf 72}, 045105 (2005). 
\bibitem{hermele}
Y. Ran, M. Hermele, P. A. Lee, and X.-G. Wen, Phys. Rev. Lett. {\bf 98}, 
117205 (2007).
\bibitem{marston}
O. Ma and J. B. Marston, Phys. Rev. Lett. {\bf 101}, 027204 (2008).
\bibitem{palee}
Y. Zhou, P. A. Lee, T.-K. Ng, and F.-C. Zhang,
Phys. Rev. Lett. {\bf 101}, 197201 (2008). 
\bibitem{lawler}
M. Lawler, A. Paramekanti, Y. B. Kim, and L. Balents,
Phys. Rev. Lett. {\bf 101}, 197202 (2008).
\bibitem{trivedi}
T. Grover, N. Trivedi, T. Senthil, and P. A. Lee, arXiv:0907.1710 (unpublished).
\bibitem{assaad}
Z. Y. Meng, T. C. Lang, S. Wessel, F. F. Assaad, and A. Muramatsu,
Nature {\bf 464}, 847 (2010).
\end{thebibliography}
\end{document}